\documentclass{aa}

\usepackage{graphicx}
\usepackage{natbib}
\usepackage{txfonts}
\bibpunct{(}{)}{;}{a}{}{,}

\def\dd{{\mathrm{d}}}
\def\ee{{\mathrm{e}}}

\def\obs#1{obs.~{\tt #1}}

\def\z{\phantom{0}}
\def\fpp{\mathit{FPP}}

\begin{document}

\title{Searching for periods in X-ray observations using Kuiper's test.}
\subtitle{Application to the ROSAT PSPC archive}
\titlerunning{Searching for periods in X-ray observations}

\author{St\'ephane Paltani\thanks{\email{Stephane.Paltani@oamp.fr}}}
\authorrunning{S. Paltani}
\institute{Laboratoire d'Astrophysique de Marseille, BP 8, 13376 Marseille cedex 12, France}

\date{Received XXX / Accepted XXX}

\abstract{We use Kuiper's test to detect periodicities in X-ray and
  gamma-ray observations. Like Rayleigh's test, it uses the individual
  photon arrival times, and is therefore well suited to the analysis
  of faint sources. Our method makes it possible to take into account
  the discontinuities in the observation, and to completely get rid of
  the contamination that results from them. This makes it particularly
  adapted to the search of periods long compared to the total
  observation duration. We propose a semi-analytical approach to
  determine the effective number of trial frequencies when searching
  for unknown periods over a frequency range. This approach can be
  easily adapted to other tests. We show that, using Kuiper's test, we
  can recover periods in frequency domains where other tests are
  completely confused by contamination.  We finally search the entire
  ROSAT Position-Sensitive Proportional Counter (PSPC) archive for
  long periods, and find 28 new periodic-source candidates.
   }

\maketitle

\section{Introduction}
\label{sec:intro}
Important efforts have been devoted to the search of periodic signals
throughout the electromagnetic spectrum.  Because of the
idiosyncrasies of astrophysical observations, different methods must
be used depending on the type of object and the wavelength range.
Four test families seem to dominate the period-detection ``market''.
The calculation of the Fourier power spectrum density
\citep[e.g.,][]{PresEtal-1993-NumRec} using a fast Fourier transform
(FFT) is adapted to evenly spaced (or evenly binned) observations.
The Lomb-Scargle periodogram
\citep{Lomb-1976-LeaSqu,Scar-1982-StuAstII,HornBali-1986-PrePer}, a
discrete Fourier transform method, can be used for unevenly-spaced
flux measurements. Epoch folding (EF) \citep[e.g.,
][]{LeahEtal-1983-SeaPul} can be used in the same conditions or for
individual photons, but requires a binning according to the phase.
Rayleigh's test \citep[e.g.,][]{GibsEtal-1982-TraEmi,Fish-1993-StaAna}
is particularly adapted for the analysis of individual photons.

Observations in the X- and gamma-rays usually have two important
characteristics. First, independent, time-tagged photons are
collected. A method requiring binning is therefore far from ideal,
as it results in a loss of information.  Furthermore, binning is
prohibited for sources detected with very few photons; for EF for
instance, the required assumption of Gaussian distribution in each bin
is not satisfied in this case. Moreover, the necessary assumptions on
the number and sizes of the bins lower the performance of the test
\citep{Schw-1999-OptPer}.  Secondly, space observations are often
interrupted by ``bad time'' periods, where no data are received.
Fourier-based methods and Rayleigh tests are seriously affected by
this problem. In practice, it means that only periods short compared
to the durations of uninterrupted observation can be investigated.

In this paper we present in detail Kuiper's test
\citep{Kuip-1960-TesCon}. This test has been applied to the
distribution of solar flares \citep{JetsEtal-1997-LonDis}, and to the
search for periodicities in Earth impacts
\citep{Jets-1997-HumSta,JetsPelt-2000-SpuPer}, but its unique
suitability to X-ray and gamma-ray observations has been overlooked.
Similarly to Rayleigh's test, it uses discrete events, and can be
applied to very faint sources without any {\em a priori} assumption.
Similarly to EF, it takes into account non-uniform coverage of the
phase domain, and can therefore be used when searching for periods
long compared to the total observation duration\footnote{In this
  paper, ``total duration'' means the time interval between the start
  and the end of an observation, including possible gaps.}.  We
study in detail the properties of Kuiper's test for period detection,
and particularly its significance level. We concentrate on two
important issues: the treatment of discontinuous observations, and the
determination of the effective number of trial frequencies when
searching for unknown periods. We finally apply the algorithm to the
entire archive of the ROSAT Position-Sensitive Proportional Counter
(PSPC) archive.

\section{Kuiper's test}
Kuiper's test \citep{Kuip-1960-TesCon} is a variant of
Kolmogorov-Smirnov's (KS) test (see \citet{PresEtal-1993-NumRec} and
\citet{JetsPelt-1996-SeaPer} for short introductions).  Given a sample
$\{x_i\}$, $i\!\!=\!\!1,...,N$, and a probability distribution
$\varphi(x)$, $a\le x\le b$, the Kuiper statistic is defined by:
\begin{equation}
  V^\Phi(\{x_i\})=\max_{a\le x\le b} \left( S^{\{x_i\}}(x)-\Phi(x) \right) +\max_{a\le x\le b} \left( \Phi(x)-S^{\{x_i\}}(x) \right),
  \label{kuiper1}
\end{equation}
where $\Phi(x)=\int_a^x~\varphi(y)~\dd y$, and
$S^{\{x_i\}}(x)=\#(x_i\le x)/N$ is the empirical cumulative
distribution of the $\{x_i\}$, $i\!\!=\!\!1,...,N$ sample ($\#(...)$
meaning ``number of $...$''). Similarly to KS, the Kuiper statistic does
not depend on the underlying distribution. The null hypothesis is that
the $\{x_i\}$, $i\!\!=\!\!1,...,N$ sample is an outcome of $N$ draws
from the $\varphi(x)$ distribution.

Kuiper's test can be readily transformed into a test of periodicity in
a series of photons by phase-folding their arrival times $\{t_i\}$,
$i\!\!=\!\!1,...,N$ for a given test period $P_0\!\!=\!\!1/f_0$:
\begin{equation}
\psi_i(f_0)=\mathrm{Frac}\left(\frac{t_i-t_0}{P_0}\right),~ i=1,...,N
\label{eq:phase}
\end{equation}
where Frac$(y)$ is the fractional part of $y$, and $t_0$ an arbitrary
time. In the absence of periodicity at frequency $f_0$, the
$\psi_i(f_0)$ phases are expected to be distributed uniformly. This
can be tested using the Kuiper statistic $V^U(\{\psi_i(f_0)\})$, where
$U(x)=x,~0\le x\le 1$ is the cumulative of a uniform distribution
between 0 and $1$.  A very low probability is evidence that the phases
are not uniformly distributed for this frequency, and indicates a
periodicity (but see Sect.~\ref{sec:contamination}).

Contrarily to KS or EF, the Kuiper statistic is invariant under a
shift of the origin for periodic distributions.  As a result,
$V^\Phi(\{\psi_i(f_0)\})$ is invariant under a shift in phase
$\{\psi_i(f_0)\}\rightarrow \{\psi_i(f_0)+\psi_0~\mathrm{mod}~1\}$
that would result from a different choice of $t_0$.

\subsection{Significance of the Kuiper statistic}
\label{sec:sig}
\citet{Kuip-1960-TesCon} gave the following asymptotic expression for
large $N$ to calculate the probability of the Kuiper statistic $V$ to
be larger than a given value $z$ under the null hypothesis:
\begin{eqnarray}
  {\bf Prob}(V\ge z/\sqrt{N})&=&\sum_{m=1}^{\infty} 2(4m^2z^2-1)\ee^{-2m^2z^2}-\nonumber\\
  &&\hspace*{-1cm}-\frac{8z}{3\sqrt{N}}\sum_{m=1}^{\infty}m^2(4m^2z^2-3)\ee^{-2m^2z^2}+O\left(\frac{1}{N}\right)
  \label{eq:asympt}
\end{eqnarray}
This is the false positive probability ($\fpp$) of falsely rejecting
the null hypothesis.  This formula is systematically used, even though
its validity for small $N$ has not been tested \citep[see,
e.g.,][]{JetsPelt-1996-SeaPer}. In Appendix~\ref{sec:uppertail} we
show that the FPP is overestimated by a factor 3 for $N\!\!=\!\!20$ at
the $10^{-7}$ level.  For $N<15$, the probability is underestimated,
which wrongly increases the rate of false positives by a factor 30 at
the $10^{-7}$ level.  Eq.~(\ref{eq:asympt}) is therefore seriously
wrong for small $N$.

\citet{Step-1965-GooFit} gives two analytical formulae valid for the
lower tail ($=1-\fpp$) of the Kuiper statistic distribution:
\begin{equation}
  \label{eq:lower-exact1}
  {\bf Prob}(V\le z)= N! \left(z-\frac{1}{N}\right)^{N-1}, \mathrm{~if~}\frac{1}{N}\le z\le \frac{2}{N},
\end{equation}
($\frac{1}{N}$ is the minimum of the Kuiper statistic), and, if
$\frac{2}{N}\le z\le \frac{3}{N}$:
\begin{equation}
  {\bf Prob}(V\le z)=\frac{(N-1)!\left(\beta^{N-1}(1-\alpha)-\alpha^{N-1}(1-\beta)\right)}
  {N^{N-2} (\beta-\alpha)}
  \label{eq:lower-exact2}
\end{equation}
with $\alpha$ and $\beta$ being the two solutions of the quadratic
equation: $t^2-(Nz-1)t+\frac{1}{2}(Nz-2)^2=0$.

\citet{Step-1965-GooFit} also gives an analytic formula for the $\fpp$:
\begin{equation}
  \label{eq:upper-exact1}
  {\bf Prob}(V\ge z)= \sum_{t=0}^M~\left(\begin{array}{c}N\\t\end{array}\right)(1-z-\frac{t}{N})^{N-t-1} T_t
\end{equation}
with:
\begin{equation}
  T_t=y^{t-3}(y^3N-y^2t\frac{3-2/N}{N}-\frac{t(t-1)(t-2)}{N^2}),
\end{equation}
where $y=z+\frac{t}{N}$, which is valid if $z\ge1/2$, if $N$ is even,
and if $z\ge (N-1)/(2N)$, if $N$ is odd.

\begin{figure}
  \resizebox{\hsize}{!}{\includegraphics{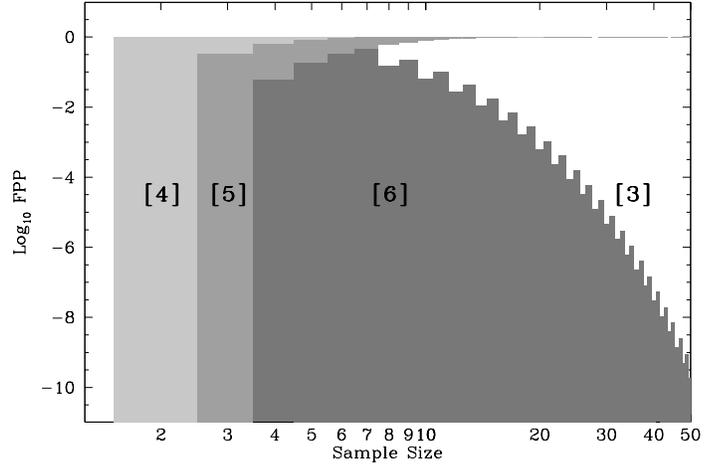}}%
  \caption{
    Domains in $\fpp$ vs Sample Size covered by the four formulae. The
    labels indicate the number of the equation used to calculate the
    probability.  The structures in the dark grey area are due to the
    different validity criteria for even and odd numbers in
    Eq.(\ref{eq:upper-exact1})}
  \label{fig:domain}
\end{figure}

The domains of validity of the three exact equations are shown in
Fig.~\ref{fig:domain}, the asymptotic formula being used outside
them. The validity condition of Eq.~(\ref{eq:upper-exact1}) is
difficult to satisfy for large $N$.  For $N\!\!=\!\!100$, the
probability that $z\ge 1/2$ is of the order of $10^{-21}$. For
$N\!\!=\!\!50$, this probability is of the order of $10^{-10}$, making
Eq.(\ref{eq:upper-exact1}) useful even for intermediate-size samples.
Eqs~(\ref{eq:lower-exact1}) and (\ref{eq:lower-exact2}) represent 40\%
of the cases for $N\!\!=\!\!10$, and only 1\% for $N\!\!=\!\!20$.
Using the four equations, the $\fpp$ is never underestimated. The only
remaining discrepancy with the true distribution is in the region
$N\sim 40-50$, where the probability is overestimated by a factor 1.5
at the $10^{-7}$ level.

\subsection{Performance of Kuiper's test}
Using extensive sets of simulations, we compare the performances of
Kuiper's test with those of the more common Rayleigh test. We create
simulated ``observations'' of periodic sources for different count
rates, different signal-to-noise ratios (S/N), and different signal
shapes.  The phase-folded light curve (hereafter simply ``light
curve'') is defined as the superimposition of a constant function (the
``continuum'') and of the first half-period of a sine function (the
``pulse''), covering a fraction $w$ of the period. The S/N is defined
as the ratio between the surfaces of the pulse and of the continuum.
We draw events at random from the ``pulse+continuum'' light curves
until a given number of events has been collected in the pulse.  For
each set of parameters, 10\,000 light curves are simulated. We then
compare the average null-hypothesis probabilities of the two tests.

\begin{figure}
  \resizebox{\hsize}{!}{\includegraphics{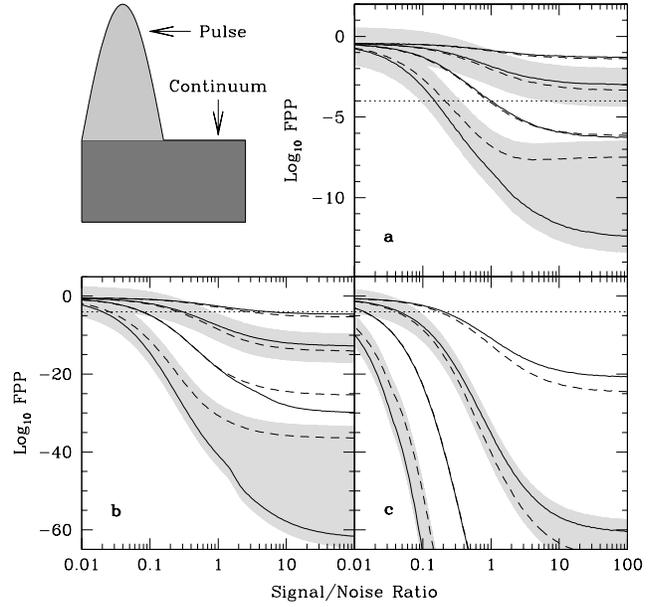}}%
  \caption{
    Sensitivities of Kuiper's and Rayleigh's tests as a function of
    the S/N between the pulse and the continuum.\textbf{(a)} 20 counts
    in the pulse; \textbf{(b)} 100 counts; \textbf{(c)} 500 counts. In
    each graph, the curves are (from top to bottom) $w\!\!=\!\!1$,
    $w\!\!=\!\!3/4$, $w\!\!=\!\!1/2$, $w\!\!=\!\!1/4$.  The
    $w\!\!=\!\!3/4$ and $w\!\!=\!\!1/4$ curves are highlighted in
    light grey for visual identification.  The solid line is Kuiper's
    test; the dashed line Rayleigh's.}
  \label{fig:perf_kuiper}
\end{figure}

Fig.~\ref{fig:perf_kuiper} shows the results for three signal
intensity cases: 20, 100, and 500 events in the pulse. In the three
cases, Rayleigh's test is more efficient for $w\ge 3/4$, while
Kuiper's performs better for $w\le 1/2$. In the situation most
favorable to Rayleigh's test (i.e.\ $w\!\!=\!\!1$, 100 events in the
pulse), the significance threshold (set arbitrarily to $10^{-4}$) is
crossed with a S/N 2 times smaller with Rayleigh's test; this
advantage decreases to 15\% with $w\!\!=\!\!3/4$, and Rayleigh's test
is about 30\% less sensitive than Kuiper's with $w\!\!=\!\!1/4$.
Kuiper's test has more difficulty with periodic signals presenting
only weak modulations, but the decrease in performance is moderate. It
is actually well known that Rayleigh's test is particularly sensitive
in the case of broad peaks \citep{LeahEtal-1983-SeaPer}. On the other
hand, some pulsars, in particular in the gamma-rays, have peaks much
narrower than those simulated here \citep{Kanb-1998-EgrObs}, in which
case Kuiper's test can significantly outperform Rayleigh's.

\section{Searching for periodicities with Kuiper's test}
\label{sec:search}

\subsection{Frequency step}

To search for periodicities, we can calculate the Kuiper statistic over a
set of test frequencies. The {\em Kuiper periodogram} (or, more
appropriately, ``frequencygram'') is defined as:
\begin{equation}
  S(f)=\log_{10} {\bf Prob}(V>V^U(\{\psi_i(f)\})),~ f_1\le f\le f_2
\end{equation}
where $V^U(\{\psi_i(f)\})$ is the Kuiper statistic calculated for a
frequency $f$. The logarithm is applied to highlight the candidate
periods. Given a periodic signal with a frequency $f_0$, Kuiper's test
may present harmonic and subharmonic peaks at frequencies $\ell\cdot
f_0$ and $f_0/\ell$ (plus their harmonics), $\ell$ being any small
integer.

To avoid missing significant peaks, $S(f)$ must be calculated for
frequencies sufficiently close to each other. Assuming a source
emitting a photon every $P_0=1/f_0$ seconds, the phases of the first
and last photons evaluated at a frequency $f_1=f_0+\Delta f$ close to
$f_0$ differ by $\Delta \varphi\simeq T\cdot\Delta f$, where $T$ is
the total duration of the observation.  The coherence is preserved if:
\begin{equation}
\label{eq:width}
  \Delta \varphi\ll 1 \Rightarrow \Delta f\ll 1/T
\end{equation}
Therefore $S(f)$ must be calculated at equidistant frequencies,
depending only on $T$. We define the oversampling parameter $k$:
\begin{equation}
\label{eq:the_width}
  \Delta f = \frac{1}{k\cdot T}
\end{equation}

Eq.(\ref{eq:width}) becomes therefore: $k \gg 1$.  If this inequality
is not satisfied, significant peaks can be missed, or underestimated
by sampling them too far from their central frequencies.  On the other
hand, the CPU time is proportional to $k$.  Reasonable values of $k$
are in the range 20--50 (but see Sect.~\ref{sec:neff}).

\subsection{Number of trials}
\label{sec:neff}
${\bf Prob}(V>V^U(\{\psi_i(f)\}))$ is the probability that $P=1/f$ is
not a period of the source {\em for a single draw of a Kuiper
  statistic}. If $S(f)$ is calculated for a set of frequencies $f_j$,
$j\!\!=\!\!1,...,n$, and assuming all the frequencies are independent,
we have:
\begin{equation}
{\bf Prob}(\exists ~j ~| V_j>z, j=1,...,n)= 1-{\bf Prob}(V\le z)^n,
\label{eq:eff_prob1}
\end{equation}
The above equation can be approximated by:
\begin{equation}
{\bf Prob}(\exists ~j ~| V_j>z, j=1,...,n)
\simeq n\cdot {\bf Prob}(V\ge z),
\label{eq:eff_prob2}
\end{equation}
under the condition $n\cdot {\bf Prob}(V\ge z)\ll 1$.  We can
therefore correct our $S(f)$ estimator for the number of trials:
\begin{equation}
\hat{S}(f) = S(f)+\log_{10} n
\label{eq:S_corr}
\end{equation}
As $n$ is proportional to $k$, Eq.~(\ref{eq:S_corr}) may destroy the
significance of some peaks if the large $k$'s required to find the
peaks are used. However, $S(f)$ is strongly correlated on scales
$\sim\Delta f$ and below, and we have ${\bf Prob}(V\ge z) \le {\bf
  Prob}(\exists ~j ~| V_j>z, j\!\!=\!\!1,...,n)\le n\cdot {\bf
  Prob}(V\ge z)$, the exact value being very difficult to calculate.
This problem affects all period search algorithms, and has been
addressed using extensive simulations for very specific cases
\citep[e.g.,][]{HornBali-1986-PrePer,DejaEtal-1988-TevTri,deJaEtal-1989-PowTes}.
We propose here a simple and workable semi-analytical method to
completely correct for the choice of $k$.

We choose an arbitrary threshold $V_*$, small enough so that $n\cdot
{\bf Prob}(V\ge V_*)\ll 1$. We then simulate $m$ sets of random
photons, and calculate $\max_{j=1,...,n} V^U({\psi_i(f_j)})$
over all $f_j$ for all $m$ simulations.  The probability that, for a
given simulation, $\max_{j=1,...,n} V^U({\psi_i(f_j)})>V_*$
can now be estimated as $\#(\max_{j=1,...,n}
V^U({\psi_i(f_j)})>V_*)/m$.  This is the left-hand side of
Eq.~(\ref{eq:eff_prob2}), with $z\!\!=\!\!V_*$. We can therefore
estimate the effective number of frequencies, $n_{\mathrm{eff}}$:
\begin{equation}
n_{\mathrm{eff}}=\frac{\#(\max_{j=1,...,n} V^U({\psi_i(f_j)})>V_*)}{m\cdot {\bf Prob}(V\ge V_*)}
\label{eq:n_eff}
\end{equation}
$n_{\mathrm{eff}}$ can be understood as the number of independent
frequencies among the $f_j$'s.  Approximating $\#(\max_{j=1,...,n}
V^U({\psi_i(f_j)})>V_*)$ with a Poisson distribution, the uncertainty
on $n_{\mathrm{eff}}$ is:
\begin{equation}
\Delta n_{\mathrm{eff}}=\frac{\sqrt{\#(\max_{j=1,...,n} V^U({\psi_i(f_j)})>V_*)}}{m\cdot {\bf Prob}(V\ge V_*)}
\label{eq:dn_eff}
\end{equation}
The corrected periodogram is then deduced from Eq.~(\ref{eq:S_corr}):
\begin{equation}
\hat{S}(f)=S(f)+\log_{10}(n_{\mathrm{eff}})
\label{eq:S_eff}
\end{equation}
Provided $\hat{S}(f)\ll 0$, $10^{\hat{S}(f)}$ is the probability that
the source has no $1/f$ period, if $n$ tests are performed.  This
method is quite general, and can be easily adapted to other
statistical tests.

In principle, the correction factor $R\!\!=\!\!n_{\mathrm{eff}}/n$ can
depend on $k$, the number of photons, the frequency range, the
observation duration, and so on, which means that $R$ should be
estimated separately for all observations.  As this is computationally
expensive for large numbers of observations (see
Sect.~\ref{sec:rosat}), we approximate $R$ as a function of $k$ only.
Details are presented in Appendix~\ref{sec:corr}.  In the limit
$k\!\!\rightarrow\!\!0$, Kuiper's tests are independent from each
other, while in the limit $k\!\!\rightarrow\!\!\infty$,
$n_{\mathrm{eff}}$ reaches a plateau. We can therefore write the
approximation:
\begin{equation}
R(k)=\frac{1}{1+r_0\cdot k}
\label{eq:R_mod}
\end{equation}

\begin{figure}
  \resizebox{\hsize}{!}{\includegraphics{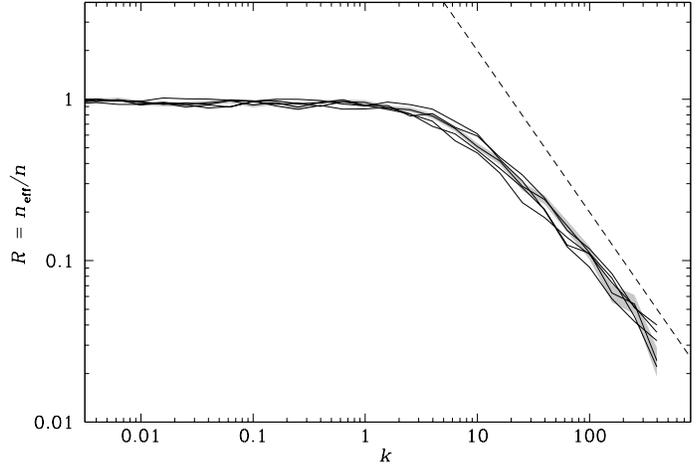}}%
  \caption{
    Ratio $R(k)\!\!=\!\!n_{\mathrm{eff}}/n$ as a function of $k$ for
    five different ``observations'': two continuous (20 and 200
    photons) ones, and three corresponding to obs.\ {\tt RP300093N00},
    {\tt RP300262N00}, and {\tt RP700232N00} with 100, 1596, and 377
    photons. The grey area shows the uncertainties for one of the five
    curves. The dashed line has a slope -1.}
  \label{fig:eff-freq}
\end{figure}

Fig.~\ref{fig:eff-freq} shows the correction factor $R(k)$ for five
sets of simulated observations with different number of photons and
different GTIs (see Sect.~\ref{sec:gti}). In all cases, 10\,000
simulations have been made for each $k$, and we set $V_*$ so that
$n\cdot {\bf Prob}(V\ge V_*)=0.1$, which seems sufficiently small.
The behavior of $R(k)$ follows quite well Eq.~(\ref{eq:R_mod}), but
the curves do significantly differ from each other, albeit moderately.
We adopt in the following a unique value $r_0\!\!=\!\!0.0815$, which
gives $n_{\mathrm{eff}}/n=0.38$ for $k\!\!=\!\!20$, or
$n_{\mathrm{eff}}/n=0.197$ for $k\!\!=\!\!50$. This value corresponds
to the upper envelope of the curves of Fig.~\ref{fig:eff-freq}.

\subsection{Discontinuous observations}
\label{sec:gti}
In high-energy observations, the photons are collected during limited
periods of time called ``good time intervals'' (GTIs).  Their main
effect is to make the cumulative distribution of the phases of the
photons coming from a constant source depart from $U(x)=x$, because
the phase intervals are not uniformly covered. This creates strong
aliases in FFTs and Rayleigh's test; EF can take into account the
actual exposure time of each phase bin, but with some limitations due
to the binning.

Kuiper's test is similar to EF in spirit, and even allows a perfect
correction for expected non-uniformity.  Like KS's test, Kuiper's test
is independent of the shape of the putative parent distribution.  Thus
we calculate exactly, for each frequency, the expected distribution
$\xi(x)$ of the phases for a constant source.  This can be done by
folding the GTIs according to the period boundaries. $\xi(x)$ being
piecewise constant, its cumulative $\Xi(x)=\int_0^x \xi(y)~\dd y$ can
be calculated exactly.  $\Phi(x)$ in Eq.~(\ref{kuiper1}) is then
replaced by $\Xi(x)$ to calculate the Kuiper statistic.

\begin{figure}
  \resizebox{\hsize}{!}{\includegraphics{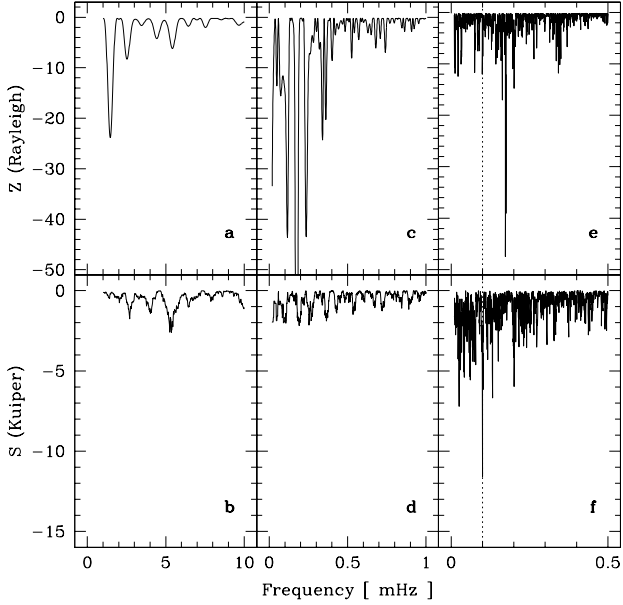}}%
  \caption{
    Effect of the GTIs on Rayleigh's (top) and Kuiper's (bottom)
    tests. {\bf (a-b)} Simulated 1000\,s 1000-photon observation.  {\bf
      (c-d)} 433-photon source in ROSAT \obs{RF500043A01} (8 GTIs).
    {\bf (e-f)} Simulated 300-photon periodic source with the GTIs of
    \obs{RP600121N00}. The dashed line indicates the location of the
    true period $P\!\!=\!\!10^4$\,s.}
  \label{fig:nonunif}
\end{figure}

Figure~\ref{fig:nonunif} compares $S(f)$ to $Z(f)=\log_{10} {\bf
  Prob}(R>R_0(f))$, $R_0(f)$ being the Rayleigh statistic, in three
different cases.
The first case is a simulated 1000\,s 1000-photon observation of a
constant source. No significant peak is observed in Kuiper's test down
to the absolute minimum frequency, $0.001$\,Hz$=\!\!1/1000$\,s, while
Rayleigh's produces several very significant spurious peaks.  The
second case is a real anonymous 433-photon source in ROSAT
\obs{RF500043A01}, an observation consisting of 8 GTIs.  Again,
absolutely no significant peak is observed in Kuiper's test down to
the absolute minimum frequency, while Rayleigh's produces many, very
deep spurious peaks.  The third case is a simulated 300-photon source
with a period $P\!\!=\!\!10^4$\,s with the GTIs of ROSAT
\obs{RP600121N00}.  The photons have been drawn from a
``$w\!\!=\!\!1/4$'' light curve with 150 photons in the pulse.  This
observation totaled $44\,733$\,s spread over 1 month in 40 separate
GTIs. The longest GTI lasted $3118$\,s, 38 of the 40 GTIs lasting half
an hour or less. The peak at $f\!\!=\!\!10^{-4}$\,Hz has comparable
depth in both tests. However, because many contaminating peaks have an
amplitude comparable to that of the true period, some even
overwhelming it, it is impossible to retrieve the $10^4$\,s period
using Rayleigh's test. In the Kuiper periodogram, the peak at
$f\!\!=\!\!10^{-4}$\,Hz dominates all other peaks with a probability
ratio larger than $20\,000$.  Furthermore, the second and third peaks
are located respectively at $f/4$ and $2\,f$, and are very probably
aliases of the true frequency.

\section{Application to known periodic sources}
\label{sec:conf}
In a search for new periodic-source candidates in the ROSAT PSPC
archive (see Sect.~\ref{sec:rosat}), we found two known periodic
sources, which particularly illustrate the power of Kuiper's test.

\subsection{EX Hya}
\begin{figure}
  \resizebox{\hsize}{!}{\includegraphics{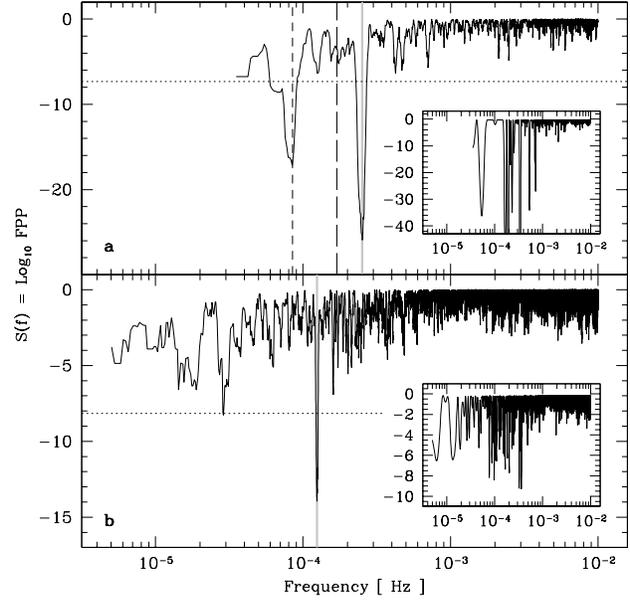}}%
  \caption{Kuiper periodograms of EX Hya {\bf (a)} and UW Pic {\bf (b)}.
    The grey lines show the $3953$\,s and $8047$\,s periods
    respectively. The dotted lines indicate the $10^{-4}$ significance
    threshold. In {\bf (a)} the short-dash and long-dash lines show
    respectively the $3\!\cdot\!3953$\,s and the possible 98-min period.
    In both panels the insets show the results of Rayleigh's test.}
  \label{fig:confirmed}
\end{figure}

\object{EX Hya} is a cataclysmic variable of type DQ Her in which a
4020\,s (67\,min) period has been claimed by
\citet{KrusEtal-1981-VarSof} using an Einstein observation. This
period was later confirmed by \citet{CordEtal-1985-ExoSof} using a
very long EXOSAT observation.  Another period of 5880\,s (98\,min) is
claimed to be present in both optical \citep{Mumf-1967-BinSta} and
X-ray \citep{CordEtal-1985-ExoSof} light curves.
Fig.~\ref{fig:confirmed}a shows the periodograms for EX Hya in ROSAT
\obs{RP300093N00}, a $28\,340$\,s observation (i.e.\ only about seven
4020\,s periods), with a $15\,542$\,s effective exposure time split in
12 GTIs.  Rayleigh's test produces a forest of spurious peaks.  On the
other hand, a very significant peak (Prob$<\!\!10^{-20}$) is easily
recovered with Kuiper's test at $P\!\!=\!\!3953$\,s, very close to the
``official'' period. The 98-min period is not found here, but there is
a second peak at about three times the 67\,min period, extremely close
(within 1\%) to $2\cdot 98$\,min. This peak could be an alias of both
periods.  The existence of the optical 98\,min period in the X-ray
domain is therefore unclear, and deserves further study.

\subsection{UW Pic}
\object{UW Pic} (RX J0531.5-4624) is a cataclysmic variable of type AM
Her with an optical period of $8010$\,s \citep{ReinEtal-1994-NewRos}.
A phase folding of the ROSAT All-Sky Survey light curve at the known
period suggests the existence of the period in the X-rays.
Fig.~\ref{fig:confirmed}b shows the Kuiper periodogram for UW Pic in ROSAT
\obs{RP300334N00}, which exhibits a very significant peak
(Prob$<10^{-6}$) at $P\!\!=\!\!8047$\,s, even though the observation
consists of 29 GTIs over 2.3 days, totalling $34\,501$\,s.  Again,
Rayleigh's test is completely unable to recover the period.

\section{Period search in the ROSAT PSPC archive}
\label{sec:rosat}
We apply Kuiper's test to the entire set of 4638 ROSAT PSPC
observations, treating them completely separately. For simplicity, we
did not attempt to combine distinct observations of a single object.
We search for periods in a range from $100$\,s up to a third of the
total duration of the observation, using $k=20$.

\subsection{Source extraction}
Source detection has been performed following the standard EXSAS
spatial analysis procedure \citep{ZimmEtal-1998-ExsUse} on a
per-observation basis using standard parameters.  Overlapping sources
were extracted twice: once ignoring the second source, and once
excluding it. We ended up with a total of 186\,572 sources, distinct
or not.  To obtain optimum sensitivity, we extracted the photons up to a
larger radius in high signal-to-noise ratios (S/N) sources than in low
S/N sources (1.5 times the source full width at half maximum compared
to 0.65 times). We extract at most $2000$ photons per source to limit
computation time.  We did not apply barycentric correction here,
because the effect is negligible for the low frequencies and
relatively short observations considered here.

\subsection{Contaminations in the PSPC observations}
\label{sec:contamination}
The ideal situation of a perfectly constant source is often not
realized in the X-ray domain. Two types of contamination affect
$S(f)$: Extrinsic periodicities, and aperiodic variability.

Spurious peaks can be produced in the Kuiper periodogram by extrinsic
phenomena.  Four different kinds of contaminations affect ROSAT PSPC
data.  One is the wobble of the ROSAT satellite: Its attitude
oscillates around the target, masking and unmasking some of the
sources behind the PSPC window support structure with a period
$\sim\!\!  400$\,s$\equiv\!\!  f_\mathrm{wob}\!\!=\!\!0.0025$\,Hz.
The spacecraft's orbit also produces contamination.  While the gaps
due to observing constraints are completely taken care of with our
method (see Sect.~\ref{sec:gti}), part of the background depends on
the position along the orbit (e.g., the scattered Solar X-ray
background \citep{SnowEtal-1994-AnaPro}), and induces a periodic
variability at the period of the spacecraft's revolution, i.e.\ 
5760\,s$\equiv\!\!  f_\mathrm{orb}\!\!=\!\!1.7361\,10^{-4}$\,Hz.  We
also found in about 50 cases a period of $86\,400$\,s, obviously of
extrinsic origin.  In a handful of observations, many objects presented
very significant peaks at $0.003$\,Hz. The fact that distinct objects
present the same period clearly indicates a non-astrophysical origin,
which we could not identify. The contaminations combine with each
other, and peaks at $f_\mathrm{wob}\pm i\,f_\mathrm{orb}$, $i$ being
any small integer, are frequent.

Knowing the contaminating frequencies, we could check whether harmonic
and subharmonic (see Sect.~\ref{sec:search}) peaks can dominate the
peak at the fundamental frequency. No subharmonic peak has been found
to dominate the fundamental, but harmonic peaks occasionally do. Thus
there is a risk of misidentifying a harmonic peak for the fundamental.

Aperiodic variability is also a serious difficulty when dealing with
long periods. When trial periods are comparable to the source's
shortest variability time scale, or longer, the effect of aperiodic
variability cannot cancel itself out over the successive phases, and
strongly affects $S(f)$, preventing period detection over large ranges
of frequencies. This is analogous to the red-noise contamination in
Fourier power spectra.

\subsection{Candidate selection}
Several thousand sources exhibit significant frequencies at the
$10^{-4}$ level (corrected for the number of trials), the vast
majority of them being due to contamination. We applied several
filters to reduce the number of candidates We rejected first all
frequencies in broad ranges around the contaminating frequencies
discussed above, and their harmonics. The ranges have been determined
using a histogram of all significant frequencies.  Aperiodic
variability has been dealt with in two steps: First, we discarded
objects for which $S(1/T_0)<-10$, $T_0$ being the total observation
duration.  We also rejected all objects for which more than 10
significant frequencies were found.  Finally, we eliminated many of
them after visual inspections, ending up with 30 objects, because
several close peaks in $S(f)$ had similar, but just below threshold,
depths.  This last step is however somewhat subjective.

\subsection{Candidate periodic sources}
\begin{figure*}
  \resizebox{\hsize}{!}{\includegraphics{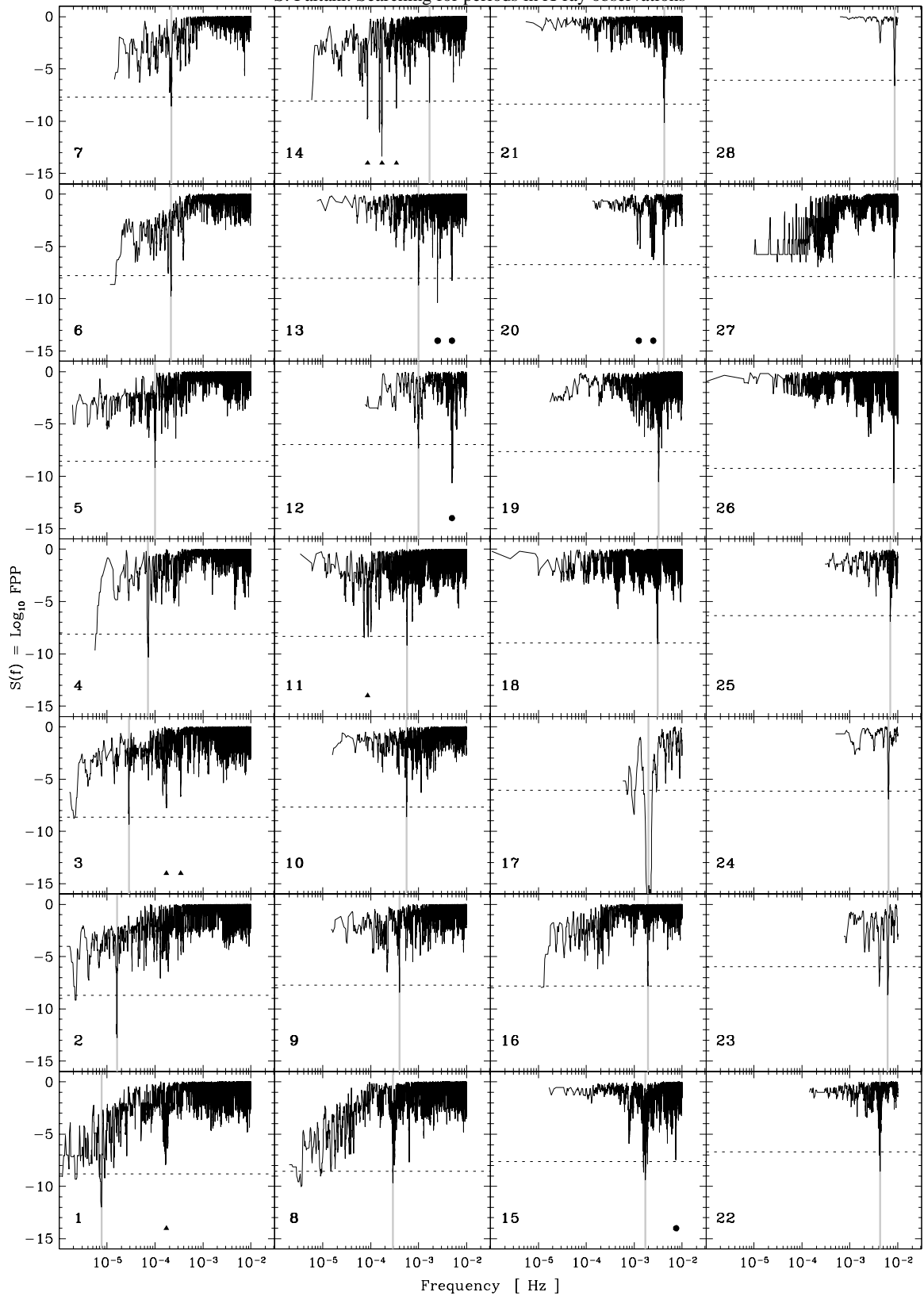}}%
  \caption{
    Kuiper periodograms of the 28 candidates listed in
    Table~\ref{tab:candidates}. The candidate frequency is highlighted
    with a grey line. The horizontal dotted line is the $10^{-4}$
    significance limit corrected for the number of trials.
    Contamination related to the wobble frequency is indicated with a
    black circle. Contamination related to the revolution frequency is
    indicated with a black triangle. The numbers are the ``ID'' column
    in Table~\ref{tab:candidates}.}
  \label{fig:pdgram}
\end{figure*}

Table~\ref{tab:candidates} lists the properties of the 30 remaining
sources.  Fig.~\ref{fig:pdgram} show $S(f)$ for the 28 new candidates.
A search over $180\,000$ objects produces about 18 spurious sources at
the $10^{-4}$ level, assuming that all contaminations are perfectly
identified. The periodicities must therefore be confirmed using
distinct data sets. Six candidates have been observed several
times using ROSAT PSPC with adequate observation durations, and are
discussed below. The 22 other sources require additional
observations before their status can be settled, and remain
candidates.

\object{V603 Aql} (Source \#17) is a classical nova for which a period
of 63 min was found using Einstein IPC data
\citep{UdalSchw-1989-PhoCat}. Using the same data,
\citet{EracEtal-1991-SeaPer} possibly find only its first harmonics,
remaining cautious about its reality.  We do not find the candidate
period in any of the two long ROSAT PSPC observations. Similarly,
\citet{BorcEtal-2003-VarOld}, combining 27 short observations, did not
find any evidence of X-ray periodicity. We found however a very
significant peak at $f\!\!\sim\!\! 0.00199$\,Hz$\equiv\!\!  503.2$\,s,
a region not explored by \citet{BorcEtal-2003-VarOld}, in
\obs{RP300262N00}.  This observation lasted 1736\,s, i.e.\ a little
more than 3 cycles.  Such a small number of cycles could result from
a chance occurrence of three similar successive flares.  However, a peak
near this frequency is found in at least two other observations, but
with a lower significance.  The repeated occurence of the peak makes
nevertheless the $503$\,s period intriguing. Its absence in most
observations could mean that it is only a characteristic variability
time scale, whitout long-term coherence, or that the periodic
modulation is not persistent.

\object{MRK 841} (Source \#20) is a Seyfert 1 galaxy, with a candidate
period of $240.68$\,s. A similar peak is found in two out of nine
other observations, which were rejected because of red noise. If
Source \#11 ($P\!\!=\!\!1741.89$\,s) is really \object{1 RXS
  J172136.9+431045}, it is also an active galactic nucleus (AGN).  AGN
do not present periodic variability in general, but, because of their
similarity to X-ray binaries, (quasi-)periodicities are not excluded.
There have been several claims of existence of periodicity in AGN
(e.g., \citet{IwasEtal-1998-DetXra} in the Seyfert 1 Galaxy
\object{IRAS 18325-5926}, but see \citet{BenlEtal-2001-QuaPer}).

Source \#22 is the symbiotic star \object{AG Dra}, and shows a
periodicity at $P\!\!=\!\!234$\,s. The peak is quite narrow, and there
is no evidence of contamination in the region surrounding the
frequency. AG Dra is a known X-ray source \citep{AndeEtal-1981-XraDet}
with two probable periods of about 350 and 550 days in the optical
\citep{FrieEtal-2003-MorPul}. No periodicity has ever been reported in
the X-rays. The period was completely absent in the few other
ROSAT PSPC observations. If real, the periodic component must be
non-persistent.

Source \#23, with a candidate period of $161.47$\,s, has been observed
three times in total. A peak at the same frequency, albeit below our
significance threshold, is found in the two other observations, making
it a very good candidate.  This object is, or is close to, the white
dwarf \object{WD 1620-391}, which appears slightly extended in the
ROSAT image. No periodicity has ever been reported for this object.

Source \#25 ($P\!\!=\!\!142.47$\,s) has been observed several times,
without any confirmation of the candidate period.

Source \#28 presents low-significance peaks around the candidate
$116.9$\,s period in other observations, but the weakness of the
source makes impossible to settle the case.

\begin{table*}
\caption{\label{tab:candidates}
  Properties of the confirmed (first two rows) and candidate periodic
  sources.  ROSAT coordinates are J2000. The frequency is expressed in mHz.
  $N$ is the number of photons. $\hat{S}(f)$ is the decimal logarithm of
  the $\fpp$ corrected for the number of trials. $N_{\mathrm F}$ is
  the number of frequencies searched. ``ID'' refers to the numbers in
  Fig.~\ref{fig:pdgram}. Identifications in italics are tentative.
  }
{
\begin{tabular}{crccccccl}
\hline
\hline
\rule{0pt}{1.2em}RA&\multicolumn{1}{c}{Dec}&Observation&Frequency&\multicolumn{1}{c}{$N$}&\multicolumn{1}{c}{$\hat{S}(f)$}&$N_{\mathrm F}$&ID&Identification\\
\hline
\rule{0pt}{1.1em}05 31 35.92&-46 24 03.8&RP300334N00&0.124324&\z125&\z-9.79&\z39715&-&UW Pic\\
12 52 24.31&-29 14 50.8&RP300093N00&0.252276&2000&-22.59&\z\z5649&-&EX Hya\\[3mm]
00 40 26.22&+30 13 57.9&RP201045N00&0.099623&\z272&\z-4.54&104776&\z5\\
03 19 01.92&+42 47 35.3&RP800035A01&0.215650&\z226&\z-5.92&\z17231&\z6\\
03 39 18.06&-45 32 57.4&RP900495N00&0.016091&1044&\z-8.04&137947&\z2\\
05 24 03.69&-00 38 51.9&RP200927N00&8.351288&\z\z77&\z-4.11&\z19846&27\\
05 28 17.51&-65 48 49.6&RP200692N00&0.289205&\z167&\z-5.07&100601&\z8\\
05 38 56.30&-64 06 01.4&RP130002N00&7.019144&1580&\z-4.50&\z\z\z639&25\\
05 40 35.01&-01 21 44.6&RP900386N00&0.071033&\z559&\z-5.20&\z35739&\z4&RX J0540.5-0121\\
08 31 53.80&+66 27 54.6&RP800022N00&0.007760&1221&\z-7.17&169674&\z1\\
10 45 04.81&-59 45 07.8&RP200108N00&8.555487&\z\z10&\z-4.49&\z\z\z335&28&\\
11 13 08.79&+20 43 38.6&RP200213N00&0.217517&\z308&\z-4.86&\z13864&\z7\\
11 20 49.83&+43 54 10.8&RP900383N00&8.272185&\z161&\z-5.39&475456&26\\
11 59 37.14&+55 46 30.8&RP700055N00&0.998413&\z609&\z-4.62&\z29978&13\\
12 06 35.38&+28 11 55.6&RP700232N00&0.568410&\z377&\z-4.94&\z12419&10\\
12 42 46.55&+31 52 04.1&RP600416N00&4.253512&\z\z96&\z-5.75&\z\z5778&21\\
12 52 01.56&+41 06 05.7&RP600050N00&1.704021&\z466&\z-4.07&\z33542&14\\
13 08 07.16&+29 06 59.0&RP110320N00&6.400603&\z\z17&\z-4.75&\z\z\z379&24\\
13 26 39.01&+30 08 53.6&RP800238N00&1.920864&\z542&\z-4.00&\z17707&16\\
13 29 50.55&+58 33 22.4&RP600458N00&3.072636&\z\z36&\z-4.08&242463&18\\
15 04 00.97&+10 26 15.2&RP700899N00&4.154825&2000&\z-4.02&\z\z1458&20&Mrk 841\\
16 01 41.65&+66 48 10.1&RP200690N00&4.262014&1354&\z-5.84&\z\z1379&22&AG Dra\\
16 06 28.24&+25 36 50.8&RP300021N00&0.399661&\z\z73&\z-4.63&\z15318&\z9\\
16 23 34.33&-39 14 14.6&RP200588A01&6.193010&\z280&\z-6.68&\z\z\z244&23&\emph{WD 1620-391}\\
16 28 33.96&+78 04 00.6&RP141845N00&0.997097&\z\z60&\z-4.29&\z\z2598&12\\
17 21 43.70&+43 11 09.5&RP300180N00&0.574087&2000&\z-4.81&\z61382&11&\emph{1RXS J172136.9+431045}\\
17 42 16.51&-29 15 08.9&RP900162N00&0.028722&2000&\z-4.66&122186&\z3\\
18 48 54.97&+00 35 02.3&RP300262N00&1.987327&1596&-17.75&\z\z\z328&17&V603 Aql\\
20 49 06.10&+30 50 08.0&RP500268N00&1.719752&2000&\z-5.72&\z11773&15\\
21 43 49.95&+38 20 43.6&RP400055N00&3.258069&2000&\z-6.87&\z11469&19\\
\hline
\end{tabular}
}\end{table*}

\section{Conclusions}
\label{sec:conc}
Kuiper's test shows very interesting properties for the search of
long-period periodic objects. Its ability to cope very naturally,
without any hidden assumption, with complex GTIs is unique.  Compared
to Rayleigh's, Kuiper's test performs better for narrow-peaked light
curves. Kuiper's test is quite sensitive to both subharmonics and
harmonics of the fundamental frequency, but usually identifies the
fundamental correctly.  Kuiper's test is particularly adapted to X-ray
missions, like XMM-Newton and Chandra, high-energy gamma-ray
satellites like GLAST, and for Cherenkov telescopes.

The semi-analytical method we propose here to correct the
false-positive probability in case of a search over a range of
frequencies should be quite useful in practice, not only for Kuiper's
test, but also for other tests, as its principle can be easily
adapted.  It has the advantage of simplicity, and of being based on
sound probability principles.

On the 28 candidate periodic sources, 6 could be cross-checked using
other ROSAT PSPC observations. Good or partial confirmation of the
existence of periodicities is found in 3 of these objects, and there
is total absence of confirmation in 3 objects. This does not
necessarily imply a ``confirmation of absence''. It must be reminded
that X-ray sources are quite often strongly variable, and that a
periodic signal may remain undetected in some observations, even
though the observing conditions seem adequate. For instance,
\citet{IsraEtal-2000-BepCha} report the detection of a periodic signal
in the X-ray pulsar \object{2E~0053.2-7242} in only one out of nine
ROSAT PSPC observations, the source having dimmed by a factor $>\!\!6$
between the different observations.

The possibility that extrinsic contamination, or statistical flukes
explain some, or even most, of the candidate periods must be
considered seriously. Firm identification of the candidates as
periodic sources will be contingent upon the detection of the periods
in independent data sets. The building up of important X-ray archives
from XMM-Newton and Chandra makes it quite probable that new observations
will be available for a fair number of these sources in the near
future.

A C library implementing the algorithms discussed in this paper is
available from the author.

\acknowledgements{This research has made use of data obtained from the
  High Energy Astrophysics Science Archive Research Center (HEASARC),
  provided by NASA's Goddard Space Flight Center. The author
  acknowledges a grant from the Swiss National Science
  Foundation}

\appendix
\section{Distribution of the Kuiper statistic}
\label{sec:uppertail}
\begin{figure}
  \resizebox{\hsize}{!}{\includegraphics{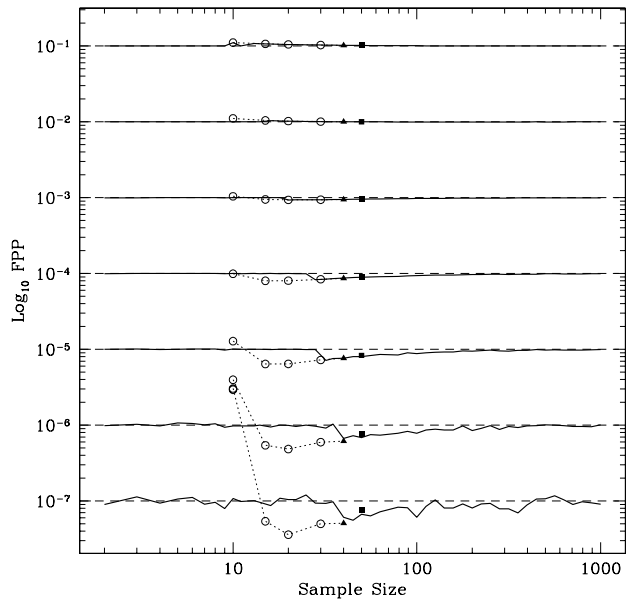}}%
  \caption{
    Fractions of false-positives compared to the expectation for 7
    different probability thresholds, from $10^{-1}$ to $10^{-7}$,
    and different sample sizes.  The empty circles have been
    calculated using Eq.~(\ref{eq:asympt}). The black symbols use
    different RNGs (see text).}
  \label{fig:uppertail}
\end{figure}

We test the formulae presented in Sect.~\ref{sec:sig} using Monte
Carlo simulations of the null hypothesis.  Fig.~\ref{fig:uppertail}
shows the fraction of test results with a probability smaller than
$10^{-1}$,..., $10^{-7}$ respectively as a function of the sample
size. These fractions would reach asymptotically $10^{-1}$,...,
$10^{-7}$ respectively if we had exact equations. $10^9$ simulations
have been performed for each sample size.  The only discrepancies are
found for sample sizes in the range 30-100.  The asymptotic formula
overestimates the probability of the null hypothesis by about 40\% for
a 40-member sample at the $10^{-7}$ level. The overestimation becomes
unimportant for sample sizes larger than 100.  An overestimation of
the FPP is however not serious, since we are chiefly concerned with
avoiding false positives.  There is no evidence of underestimation of
the FPP, which would be a more serious issue, as it would lead to
false negatives.

Empty circles in Fig.~\ref{fig:uppertail} have been calculated using
Eq.~(\ref{eq:asympt}) only. The overestimation reaches a factor 3 for
$N\!\!=\!\!20$. More importantly, the FPPs are underestimated for
$N\!\!<\!\!15$. The factor reaches 30 for $N\!\!=\!\!10$ at the
$10^{-7}$ level, and is close to 300 at the $10^{-8}$ level.

In principle, one should be cautious about simulations exploring tails
of probability distributions, since the random number generators
(RNGs) may present defects in these regimes.  This does not seem to
be a problem here. Indeed, the simulated FPPs match perfectly the
expected ones when either an exact formula is used, or when $N$ is
large enough if Eq.~(\ref{eq:asympt}) is used.  Moreover, the simulations
that end up in the very tail of the Kuiper-statistic distribution are
not at all in the tails of the uniform distributions used to generate
the list of photons.

We further checked the validity of the simulations by comparing
different RNGs. The curves used the MT19937 generator
\citep{MatsNish-1998-AcmTra}. The black triangles at $N\!\!=\!\!40$
used the RANLUX generator at luxury level 2 \citep{Lusc-1994-PorHig},
and the black squares at $N\!\!=\!\!50$ used the (very poor) standard
UNIX RNG (C function {\tt rand()})\footnote{These algorithms can be
  found in the GNU Scientific Library at {\tt
    http://sources.redhat.com/gsl/}}.  The different RNGs produce
perfectly compatible results within the statistical fluctuations due
to the limited number of simulations, which reach $10\,\%$ at the
$10^{-7}$ level, and $1\,\%$ at the $10^{-5}$ level.

\section{Details on the $R$ correction factor}
\label{sec:corr}
\begin{figure}
  \resizebox{\hsize}{!}{\includegraphics{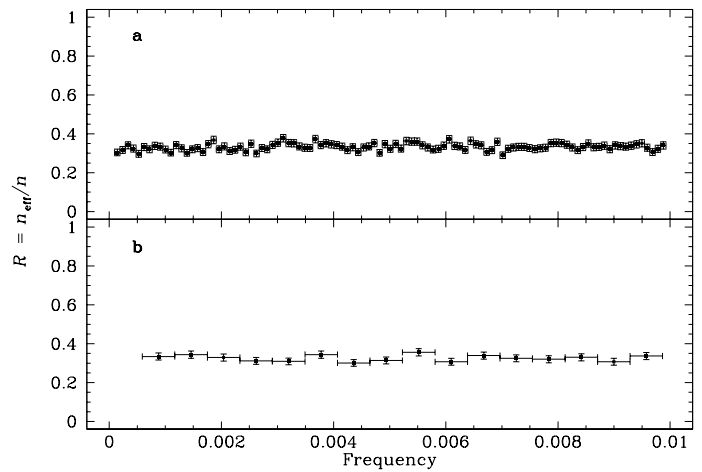}}%
  \caption{
    Correction factor $R$ for successive chunks of 1000 frequencies
    for two simulated 100-photon observations using the GTIs of
    \obs{RP201045N00} {\bf (a)} and {\tt RP800035A01} {\bf (b)}.}
  \label{fig:range}
\end{figure}

We explore in more detail the properties of the correction factor $R$
with respect to other parameters. Fig.~\ref{fig:range} shows the
effect of the frequency range. We cut the set of trial frequencies
into chunks of 1000 frequencies for two 100-photon simulated
observations covering the GTIs of obs.~{\tt RP201045N00} and {\tt
  RP800035A01}. We used $k=20$, and performed 10\,000 simulations in
each case.  While not constant, $R$ changes moderately, without any
visible trend.

\begin{figure}
  \resizebox{\hsize}{!}{\includegraphics{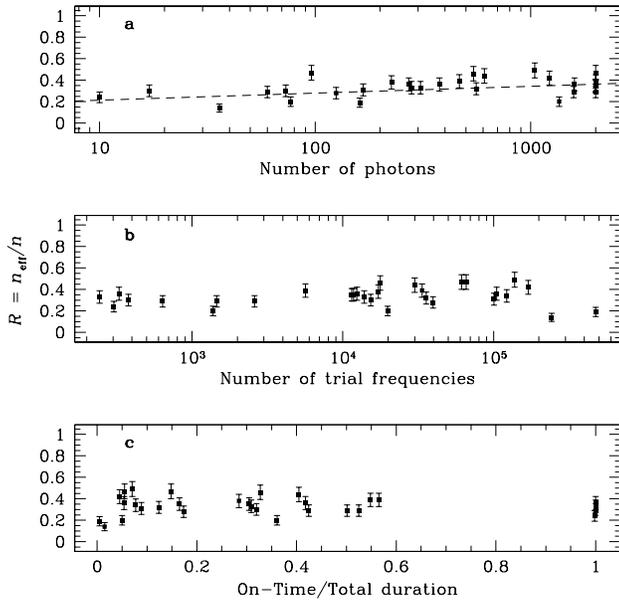}}%
  \caption{
    Correction factor $R$ for the 30 sources from
    Table~\ref{tab:candidates} as a function of $N$ {\bf (a)},
    $N_{\mathrm F}$ {\bf (b)}, and the on-time vs total duration ratio
    {\bf (c)}. The dashed line is the best linear fit.}
  \label{fig:depend}
\end{figure}

We test the dependence of $R$ on the number of photons $N$, the number
of trial frequencies $N_{\mathrm F}$ (which is roughly proportional to
the observation duration), and a measure of the importance of gaps in
the observation, given by the ratio between the ``on-time'' (i.e.\ the
sum of the individual GTI durations) and the total duration.
Fig.~\ref{fig:depend}
shows $R$ for the 30 sources from Table~\ref{tab:candidates} as a
function of $N$, $N_{\mathrm F}$, and on-time vs total duration ratio
for $k=20$, with 1000 simulations per observation. A significant
correlation is found only between $R$ and $N$, making $R$ increase
with $N$, with a 1\% probability chance occurrence of Spearman's
correlation coefficient.

The average $R$ value for $k=20$ is $0.335$, with a rms corrected for
the contribution of the number of simulations of $0.066$.  After
removal of the best linear fit to Fig.~\ref{fig:depend}a, the rms
becomes $0.053$. Hence most of the scatter remains unexplained, and
probably results from the distribution of the GTIs.

All parameters except $k$ can be neglected as a first approximation,
and using $R(k)$ is justified. A more detailed approximation would
make use of both $k$ and $N$. Unless one is searching for periodic
sources in a large number of observations, which is the case in this
work, the correct approach is nevertheless to calculate $R$
specifically for the observation at hand.

\bibliography{biblio}
\bibliographystyle{apj}

\end{document}